# Femtosecond laser direct-write photoconductive patterns on tellurite glass

Gözden TORUN[1,*], Anastasia ROMASHKINA[1], Tetsuo KISHI[2], and Yves BELLOUARD[1]

Gözden TORUN*, Anastasia ROMASHKINA and Yves BELLOUARD
Galatea Laboratory, STI/IEM, Ecole Polytechnique Fédérale de Lausanne (EPFL), 2002 Neuchâtel, Switzerland
E-mail: gozden.torun@epfl.ch

Tetsuo KISHI
Department of Materials Science and Engineering, Tokyo Institute of Technology, 152-8552, Tokyo, Japan

We report the formation of arbitrary photoconductive patterns made of tellurium (Te) nanocrystals by exposing a tellurite ($TeO_2$-based) glass to femtosecond laser pulses. During this process, Te/$TeO_2$-glass nanocomposite interfaces with photoconductive properties form on the tellurite glass substrate. We show that these laser-written patterns have a highly reproducible photo-response, from the near ultraviolet (263 nm) to the visible spectrum, stable over a few months. Specifically, high responsivity (16.55 A/W) and detectivity ($5.25\cdot10^{11}$ Jones) of a single laser-written line pattern are measured for an illumination dose of 0.07 $mW/cm^2$ at 400 nm. This work illustrates a pathway for locally turning a tellurite glass into functional photoconductor of arbitrary shape, without adding materials and using a single laser process step.

## 1. Introduction

Recently, we demonstrated that upon near-IR femtosecond (fs) laser exposure, the tellurite ($TeO_2$-based) glass evolves into a semiconductor/glass composite[1] consisting of trigonal tellurium (t-Te) nanocrystals embedded in a $TeO_2$ glass matrix. Extensive investigation revealed that the laser-modified zones show evidence of t-Te nanoparticles and nanocrystals with a low amount of amorphous Te (a-Te) upon a single femtosecond laser pulse exposure of $TeO_2$-based glass[2]. In particular, it leads to the formation Te/$TeO_2$-glass interface at the surface, at which the ratio of phases depends on the laser processing parameters. This process is based on a scalable laser direct-write technology by focusing a femtosecond laser on the surface of a glass substrate and scanning the laser spot as a pattern with an arbitrary length and shape.

Previously, several authors reported the production of $TeO_2$/Te interface by surface oxidation of pure tellurium thin films[3–6]. In these studies, the oxidation process was driven by ultra-violet (UV) irradiation or by continuous-wave (CW) lasers operating at 440–520 nm[7], and several functional properties, e.g., photoconductivity[3,4,8,9], ultrahigh chemical sensitivity[10,11], and better optical properties than Te and $TeO_2$[12], were demonstrated. Here, we follow a different path. Starting from glass, a $TeO_2$-based transparent substrate of arbitrary thickness, we use a femtosecond laser pulse to *transform* it into a pure t-Te phase locally and with micron-scale resolution. Apart from its inherent versatility since any patterns can be produced on any substrate form and size, this approach has the potential to be simple and economical. Thanks to a single process, *the functional device is produced by transforming locally a plain substrate without adding any other materials.*

In the sequel, we demonstrate this powerful concept by investigating the photoconductive properties of the Te/$TeO_2$-glass interface produced by direct-write femtosecond laser exposure. Specifically, we unravel a highly reproducible and sensitive photo-response under different illumination conditions in the near ultraviolet to the visible (UV-VIS) spectrum that we characterize for various laser exposure parameters and illumination conditions. This single process is particularly appealing for light-sensing devices of arbitrary sizes and shapes made by functionalizing a single piece of material.

## 2. Experimental Section/Methods

### *Glass substrate preparation and femtosecond laser micromachining*

The glass composition tested in this study is $10K_2O$-$10WO_3$-$80TeO_2$ (mol%). Commercial powders were mixed and melted in an Au crucible at around 973 K for 30 min in an electric furnace to produce the substrates. The melt was then quenched onto a brass plate. After quenching, the sample was crushed and remelted at 973 K for 30 min, followed by subsequent annealing at 598 K for 1 hour. The so-obtained glass rods were cut and optically polished to samples with a thickness of 1 mm for further optical characterization and femtosecond laser machining.

An Yb-doped femtosecond fiber laser (Yuzu from Amplitude) emitting 270 fs pulses at 1030 nm was used in this experiment. Laser patterns consisting of parallel lines with several lengths from 1 to 10 mm were inscribed on the surface of the tellurite glass. The specimen was translated under the laser focus using high-precision motorized stages (Ultra-HR from PI Micos). The laser beam was focused on the surface of the sample using a 0.4 numerical aperture (NA) objective (OFR-20x-1064 nm from Thorlabs), resulting in a spot size (defined at $1/e^2$) of ~1.97 µm. The repetition rate was fixed at 1 MHz, corresponding to a thermal cumulative regime for tellurite glass[1]. Here, the number of effective pulses per spot varies from 20 to 4000, and the range of pulse energy spans from 1 to 200 nJ, resulting in an incoming net fluence ranging from 0.0066 to 263 $J/mm^2$. Further details on the exposure parameters can be found in our previous study[2].

### *Sample characterization*

After laser exposure, the tellurite glass samples were first observed using an optical microscope (OM, BX51 from Olympus). A Raman spectrometer (LabRam HR from Horiba), equipped with a 532 nm laser excitation source attenuated down to 4 mW, focused with a 0.9 NA objective (100x-532 nm from Thorlabs) down to a micron-size spot was then used to confirm the presence of elemental Te nanocrystals in laser-modified zones. Further, another Raman spectrometer (MonoVista CRS+ from Spectroscopy & Imaging GmbH), equipped with a 442 nm laser-excitation source (He-Cd laser from Kimmon Koha) with an incident power of 115 mW was used to investigate the degradation mechanism under UV-light irradiation. The linearly-polarized Raman laser beam was focused at the surface of the glass sample using a 0.9 NA objective (50x-532 nm from Thorlabs). A series of point scans were performed with acquisition times of 60 s per individual spot. Finally, the absorption spectra were measured at room temperature for wavelengths ranging from 250 to 2500 nm using an ultraviolet-visible-near-infrared spectrometer ((UV-VIS-NIR, Lambda 950 from Perkin Elmer). For this purpose and to isolate the functionalized regions in the substrate, a mask with a hole of around 2x2 $mm^2$ was cut out of a paper sheet for broadband absorbance. For the measurement, the reference beam power was attenuated to 10% to compensate for the presence of the mask and to ensure the effective reduction of the beam size from the original 2 cm in diameter. The thickness of the sample used for this transmission measurement was 2 mm.

***Electrical measurement, device fabrication and characterization***

First, the DC resistivity of the laser-written tracks was measured by a four-probe station equipped with a microscope connected to a source measurement unit (SMU B2902A from Keysight) applying a bias voltage of 40 V. A control software (Quick I/V Measurement Software from Keysight) was used to obtain the data. The tungsten probes were placed on the sample with the help of high-precision adjustments. The conductivity temperature dependence was measured by slowly heating the sample from room temperature to 368 K using a heating stage (from Linkam).

To fabricate the device used for probing the photoconductivity, thin gold electrodes (~20 nm-thick) were sputtered (JFC-1200 Fine Coater from JEOL). For obtaining the desired shape of electrodes, a hard mask made from fused silica glass by femtosecond laser machining-assisted etching process was directly placed on the tellurite glass before sputtering, shown in **Figure S1**. Later, wire bonding (HB10 wedge and ball bonder from TPT) with either Au or Al wires was used to interface the glass laser-written photoconductive wire with a standard printed circuit board (PCB). To ensure the wires were attached to the sample and the PCB over long measurement periods, the electrodes were covered with Ag-based electrically conductive epoxy (H20E-FC from epoxy technology; the resistivity is less than 0.04 Ω·m at room temperature). The epoxy was later cured at 393 K for 15 minutes.

After the device fabrication, characteristic transient current-voltage responses were collected with the source measurement unit (SMU B2902A from Keysight). Various light sources were used for photoconductivity measurements: a white-light LED (spectral range spanning from 450 to 750 nm) and three other LEDs with central wavelengths at 460 nm, 400 nm, and 263 nm emitting various optical intensities (up to 2.1 mW/cm$^2$). The spectral profiles of the LED sources are presented in **Figure S2**. After collimation, a cylindrical lens was used to create a stretched elliptical profile illumination covering the laser-written patterns with an area of approximately 200 mm$^2$. The same measurement was performed for at least ten different lines for statistical purposes and over a few months to observe material degradation. Here, one month refers to 30 days of consecutive measurements. **Figure 1** illustrates a schematic representation of the device fabrication and the characterization procedure.

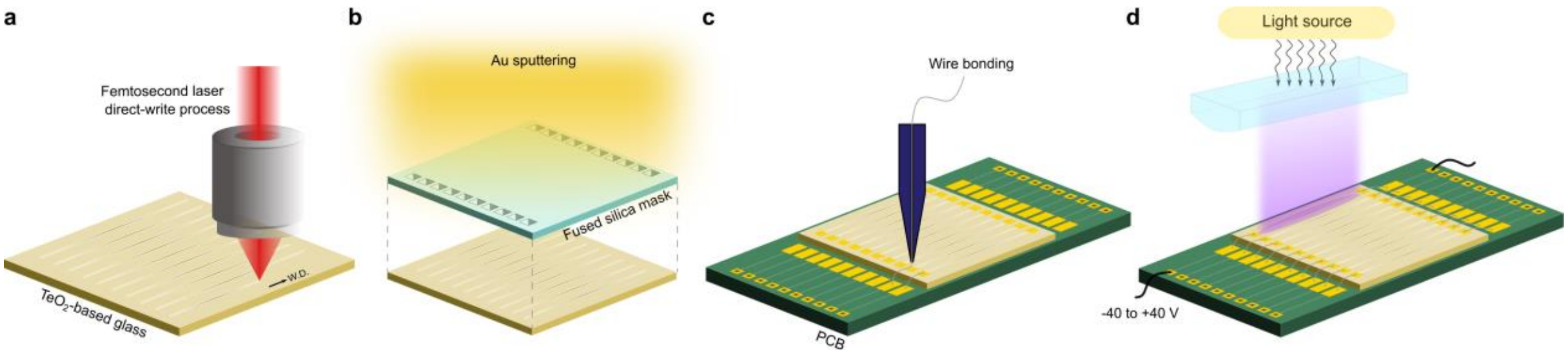

**Figure 1:** Schematic representations of the device preparation steps a-c) and photoconductivity characterization method d). a) Femtosecond laser direct-write process applied to tellurite glass: the line patterns were produced by moving the specimen under the laser beam at a prescribed velocity. b) The fabrication of Au electrodes at both ends of the sample was performed by sputtering the metal through a hard mask to protect the central portion. c) Wire bonding was used to connect the electrodes on the specimen to a standard PCB on which the glass substrate was mounted. d) Device characterization was performed by applying a bias voltage ranging from -40 to +40 V through the PCB pins. The illumination source for photoconductivity characterization was delivered in the form of a thin elliptical spot exposing individual line patterns separately.

## 3. Results and discussion

### *2.1. Electrical properties*

The electrical conductivity of pure $TeO_2$ glass and binary $TeO_2$ glass systems are based on a small-radius polaron hopping mechanism at and above room temperature[13]. Depending on the glass modifier oxides, there is an additional contribution from ionic conductivity. The composition studied in this work ($10K_2O$-$10WO_3$-$80TeO_2$ (mol%)) has a reported electrical resistivity (DC) at room temperature of $1.4 \cdot 10^{18}$ $\Omega \cdot$m[14].

**Figure 2** illustrates the effect of laser processing parameters and temperature on the DC electrical resistivity, absorption spectra, and direct and indirect band gap plots of both pristine and laser-written patterns. The resistivity of the patterns was calculated from the measured resistance and the effective dimensions (i.e., cross-section and length) measured from both the images taken by OM and transmission electron microscopy (TEM)[2], respectively. The resistivity of the patterns decays from ~50 Ω·m to ~0.001 Ω·m with the increasing number of pulses (Figure 2a). Several parameters, such as the surface area of grain boundaries, preferred grain orientation, presence of impurities, crystallographic defects (vacancies), and other structural defects, influence DC resistivity[15]. We showed that the laser-modified zone consists of Te nanocrystals (with a grain size range of ~5-15 nm) growing proportionally with the number of laser pulses and the laser electric field intensity[2]. Low laser fluence results in thinner crystallized areas with disconnected nanocrystals, more susceptible to charges, impurities at the interface, and surface scattering. Through the grain-growth mechanism stimulated by higher laser fluence in the thermal-cumulative regime, the value reaches the resistivity of polycrystalline bulk Te[16–18]. However, no correlation was observed between the orientation of the electric field and the dark resistivity of line patterns. All patterns tested from this point were written at a pulse energy of 200 nJ and with 4000 pulses per focal spot (corresponding to an incoming pulse fluence of 262 J/mm$^2$).

Figure 2b shows the temperature-dependent relative dark resistivity of the laser-written lines, reflecting a typical semiconductor behavior due to the thermally-active charge transportation. The inset shows the activation energy, estimated to be ~0.6 meV using an Arrhenius-law fitting procedure (ln $R_T$ vs. $10^3/T$)[15,19,20]. Similarly, the resistivity of the pristine glass drops with temperature ($1.14 \cdot 10^{14}$ Ω·m at 373 K[14]).

Figure 2c presents the absorption spectra of the *pristine glass* versus the *laser-written area* of 2x2 mm$^2$. The laser-written area shows broadband absorption in the solar spectral emission range, decreasing continuously towards the band gap energy of t-Te (~3600 nm). While it is above 80 % in the visible spectrum, there is a sharp rise in absorption below 460 nm, reaching the maximum value between 330-410 nm. Two broad peaks in the absorption curve are also observed, at ~400 nm and ~1000 nm, respectively, similar to previously reported results[21].

Theoretical and experimental studies related to the optical absorption of pure Te indicate the presence of two peaks: one in the range of ~3-6 eV (i.e., ~400-200 nm), associated with the direct transition from the valence band (p-bonding triplet) to the conduction band (p-antibonding triplet), and another one in the range of ~0-3 eV (i.e., from the mid-IR to 400 nm), assigned to forbidden direct-transition from the valence band (p-lone pair) to the conduction band (p-antibonding triplet). In addition, the absorption spectrum of Te-nanorods exhibits some differences when compared to that of bulk Te, presenting a broad absorption peak due to the allowed direct transition located at approximately 300 nm[22]. For Te nanoparticles, a broad absorption band at around 660 nm indicates a forbidden direct transition[23]. In our case, the peak around 330-410 nm is corresponding to the direct band gap transition, while the forbidden direct transition is located at ~600-1000 nm. In another study, Te nanoparticles between 10 nm and 120 nm show a plasmonic-like resonance-dominated transition in the spectral range ~300 to 400 nm[24]. It is not completely a plasmonic resonance due to the absence of ground-state free carriers as it is in metals or doped semiconductors. Above 120 nm, Te nanoparticles exhibit all-dielectric (Mie-type) resonance. Similar to our study in which the sizes of nanocrystals and nanoparticles are distributed between ~5 and 55 nm, the absorption spectra of Te nanoparticles (with sizes ranging from 10 nm to 300 nm) cover the entire solar emission spectrum from 300 nm to 2000 nm[24]. The presence of localized states is due to interfaces at the surface, the grain boundary, and intra-grain regions. The impurities further absorb the photon energy below the band gap through one of the gap states within defect density, contributing to broadband absorption. The absorption spectra can be modulated by the laser-writing parameters, as shown in **Figure S3**.

Figure 2 shows Tauc plots[25] of both direct and indirect bandgap model behaviors. While a linear region (and hence, a bandgap value) is identifiable for the non-exposed case, the laser-exposed regions exhibit less pronounced linear zones that may account for their disorganized structures and interfacial effects. A shift towards lower energy, both in the direct and indirect optical band gap models, is observed as an expected consequence of the presence of Te-nanocrystals after laser exposure.

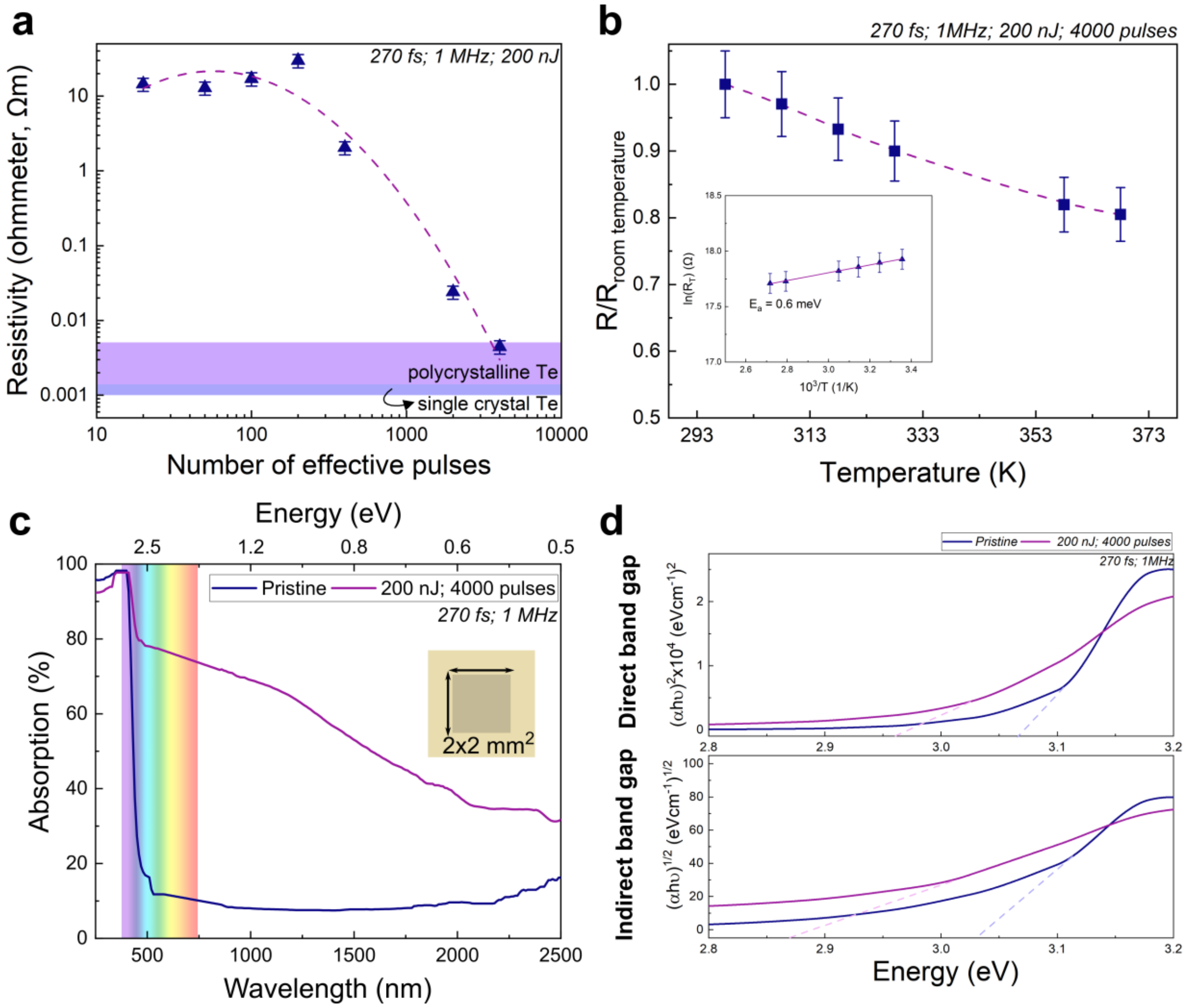

**Figure 2:** a) Effect of the laser writing parameters on electrical (DC) resistivity of line patterns. b) Effect of the temperature on the resistivity of laser-written patterns with the inset image of the Arrhenius-law fitting. c) Measured UV-VIS-NIR absorption spectra of the pristine and laser-written area (2x2 $mm^2$), d) Corresponding Tauc plots of pristine and the laser-written area, considering both indirect and direct bandgap absorption models.

### *2.2. Photoconductive properties and long-term stability*

The p-type behavior of Te originates from the p-state lone-pair electrons that form the upper level of the valence band[26]. The population of holes in p-type conduction comes from the interaction of these lone-pair electrons with dangling bonds. A region enriched with holes forms at interfaces such as the surface, the grain boundary, and intra-grain regions due to vacancies or impurities. The photoconductivity can occur at the interface due to the variation in hole concentration[11]. In addition, there can be an adsorption process, i.e., gas atoms adsorbed during the measurement[4,5,11] affecting the population of holes. In another Te/$TeO_2$ system, photoconduction is explained by the presence of photo-excited carriers in the Te-layer, which are trapped at the interface of the $Te/TeO_2$ layer[4,8]. In our case, the laser-modified zone consists of a mixture of polycrystalline Te (t-Te nanoparticles and nanocrystals), amorphous Te (a-Te), and $TeO_2$-glass[2], forming a heterogeneous assembly. The photoconductivity mechanism, hence, is likely more complicated than in well-defined $Te/TeO_2$ interfaces[3,4,8].

To avoid manual manipulations while testing the photoconductivity of the laser-written patterns, we designed a specific device in coplanar configuration, illustrated in **Figure 1d**. The electrodes in the form of a thin film of Au at both ends of the line patterns are connected to a printed circuit board (PCB) by wire bonding. We conducted a current-voltage (I-V) characterization and calculated their responsivity, detectivity, external quantum efficiency, and generated photocurrent by varying optical power density. Additionally, we determined the spectral and temporal photo-responses of the patterns. The results are gathered in **Figure 3**.

The spectral response of the $Te/TeO_2$-glass nanocomposite is demonstrated in Figure 3a. It exhibits a strong absorption at ~400 nm, leading to the highest photocurrent output. The spectrum of the white light is presented in **Figure S2**. Although the center peak is located at ~600 nm, radiation below 460 nm can contribute to the observed photo-response under white light illumination. I-V curves obtained without and with illumination in both forward and reverse bias are displayed in Figure 3b. The typical I-V curve is deviating from ohmic contact in the range of -40 V to +40 V. The absolute semi-log I-V curve is presented in **Figure S4**. In the dark current for a zero bias voltage, we measure near-zero current ($3.7\cdot10^{-10}\pm1\cdot10^{-10}$A). The electrical conductivity in the device is controlled mainly by the conductive part of the laser-modified zones, interfaces, and structural impurities and defects[27]. For the current to flow, the device in this scheme requires a bias voltage above zero. Once the bias voltage is exceeded the internal barrier voltage (around 1 V in this case) and the knee voltage (or breakdown voltage for reverse bias) is surpassed, the external electric field supplies extra carriers. By illuminating the active area of the device, the current rises as a function of optical power density. The photocurrent at 400 nm versus optical power density follows a typical square root-like relation, typically observed in the case of high resistances.

Figure 3c shows the temporal response of laser-written patterns under the illumination of 400 nm with different optical power densities. From this graph, we calculated the responsivity ($R_{ph}$), detectivity ($D^*$), and external quantum efficiency (EQE), which are a figure of merit (FOM) for the photo-detecting properties of our laser-written patterns on tellurite glass. R indicates the generated photocurrent per unit area, D* displays the ability to differentiate weak signals from noise, and EQE the number of charge carriers (electron-hole pairs) collected per photon incident on the photodetector[28]. They are expressed as:

$$R_{ph} = \frac{\Delta I}{AP}, \text{ where } \Delta I = I_{ph} = I_{light} - I_{dark}$$

$$D^* = \frac{R_{ph}\sqrt{A}}{\sqrt{2eI_{dark}}}$$

$$EQE\ (\%) = \left(\frac{hcR_{ph}}{e\lambda}\right) \approx \left[\frac{R_{ph}}{\lambda}\right] \times 1240\ (W\frac{nm}{A})$$

where $I_{ph}$, $I_{light}$, $I_{dark}$, P, A, e, h, c, λ are the generated photocurrent, the current measured under illumination, the dark current, optical power density, the effective covered area with Te on the $TeO_2$-based glass, the charge of an electron, the Planck constant, the speed of light and the illumination source wavelength, respectively. At 400 nm and under an exposure dose of 0.07 $mW/cm^2$, the peak responsivity and detectivity of the device are calculated to be ~16.55 A/W and $5.25 \cdot 10^{11}$ Jones, respectively. The responsivity is reduced with the optical power density, a behavior similar to various Te photodetectors[4,8,19,22,29–31], which is attributed to defect states in the laser-modified zones. Photo-generated holes under low optical intensity are captured by the defect states near the valence band and reduce the number of recombination of electron-hole pairs. However, under high optical intensity, a low number of photo-generated holes are captured, due to the limited number of defect states. Therefore, the laser-modified zone is more sensitive under lower light intensities. A similar trend is observed in the EQE (%) plot, which is directly proportional to the responsivity and inversely proportional to the light intensity. Note that the calculated responsivity, detectivity and EQE indicate the peak values, which can alternate due to geometrical uncertainties, i.e., fluctuation in the active part of the laser-modified width, connectivity and homogeneity of nanocrystals in the laser-modified zone. The generated photocurrent follows an empiric power law in the form:

$$I_{ph} = \beta P^{\alpha}$$

where α is a dimensionless exponent ($\leq 1$), providing information related to the number of traps (or defect states) present, and β is a parameter related to the photodetector responsivity[32]. α equals 1 in an ideal trap-free photodetector but becomes less than 1 in the presence of trap states. In our case, α is 0.46, implying that most traps are already filled at lower optical power densities, and additional illumination power rises the photocurrent less efficiently. The internal quantum efficiency (IQE), the ratio of the number of charge carriers or electron-hole pairs generated to the number of photons absorbed[33], is 5012 % for the same illumination conditions (with 97.73 % absorption at 400 nm). There are a few possible reasons why a high absolute responsivity and high EQE (i.e. >100%) are measured. The higher external bias voltage above the barrier voltage leads to the generation of additional 'non-light driven' electron-hole pairs. At the illumination wavelength, which is much higher than the bandgap of Te (0.34 eV[34] in homogeneous bulk tellurium, but can be engineered up to ~1.42 eV by modulating its size at nanoscales[35]), more electron-hole pairs can be stimulated by avalanche multiplication per photon in the active region of the device. In addition, trapped minority carriers, electrons in our case, at the various defective states in our system, further enhance the internal gain. Although more than 100% EQE is not common, a high internal gain is possible by avalanche carrier multiplication[36], such as by metal-semiconductor-metal (MSM) photodetectors or p-i-n photodiodes[37].

The temporal evolution of the Te/$TeO_2$-glass nanocomposite is displayed in Figure 3e. The rise time, the time required for current to increase from 10 % to 90 % of its peak value under illumination, is one of the key parameters to evaluate a photodetector performance. We measured an average rise time of about 20 s, which is rather low. The carrier mobility in Te is temperature-dependent and typically ranges from 20 to 50 $cm^2$/Vs at room temperature[38]. The mobility is also affected by the contribution of carriers present near the Te/$TeO_2$ interface[22]. Therefore, the delay in the rise time of the Te/$TeO_2$-glass interface is attributed to several factors, such as the existence of charge impurities, defects, or trap states. A decrease in the mobility of carriers can be also due to surface texture originating from the laser-induced self-organized nanostructures[1]. In addition, high photoconductive gain results in an excessive number of carriers, generally referring to a photo-multiplication[39], which is a slow process because the photocurrent generation rate is higher than the recombination rate in our case during the rise time. In contrast, the decay time, which is the time required for the current to decrease from 90 % to 10 % of its peak value, is an order of magnitude slower. Likewise, a slow photoconductive decay response can be attributed to a large number of recombination centers, the presence of many trap levels, and defect states within the band gap as such that recombination of minority carriers takes a longer time[40]. Relaxation curves are best fitted with a sum of two exponential functions, expressed as:

$$I_{ph}(t) = I_{dark} + A_1 e^{-\frac{t}{\tau_1}} + A_2 e^{-\frac{t}{\tau_2}}$$

where $A_1$ and $A_2$ are weight coefficients, and $\tau_1$ and $\tau_2$ are the characteristic decay constants (effective relaxation times)[41,42]. While the average rise time is approximately 20 s, the decay time is about an order of magnitude longer[43]. This persistent photoconductivity is also observed in amorphous, metal-oxide semiconductors and wide-bandgap semiconductors[44,45]. Various reasons for the persistent photoconductivity are the presence of oxygen vacancies located within the bandgap[44], large spatial fluctuations of the potential energy of charge carriers[45], point defects, and metastable defects[46]. In some cases, the persistent behavior can take a few hours to days without illumination. This behavior can be eliminated by designing a device with a three-gated terminal in a sandwich configuration, such as a field-effect transistor device structure, and subsequently by applying a short pulse positive gate voltage[44,47]. The persistent photoconduction indicates that there is a complex recombination process of free carriers by both radiative and non-radiative in the Te/$TeO_2$-glass interface. For different wavelengths, **Figure S5** shows the spectro-temporal evolution of the line patterns. The rise time is 82 s for white light, 55 s for 463 nm, and 41 s for 263 nm. The decay times for white light, 463 nm and 263 nm under the same illumination conditions are 411 s, 207 s, and 286 s, respectively. Various Te-based photodetector (whether macro- or nano-scale, from ultraviolet to near-infrared) shows the rise and decay times in the range of microseconds to a few hours[4,8,19,22,29–31].

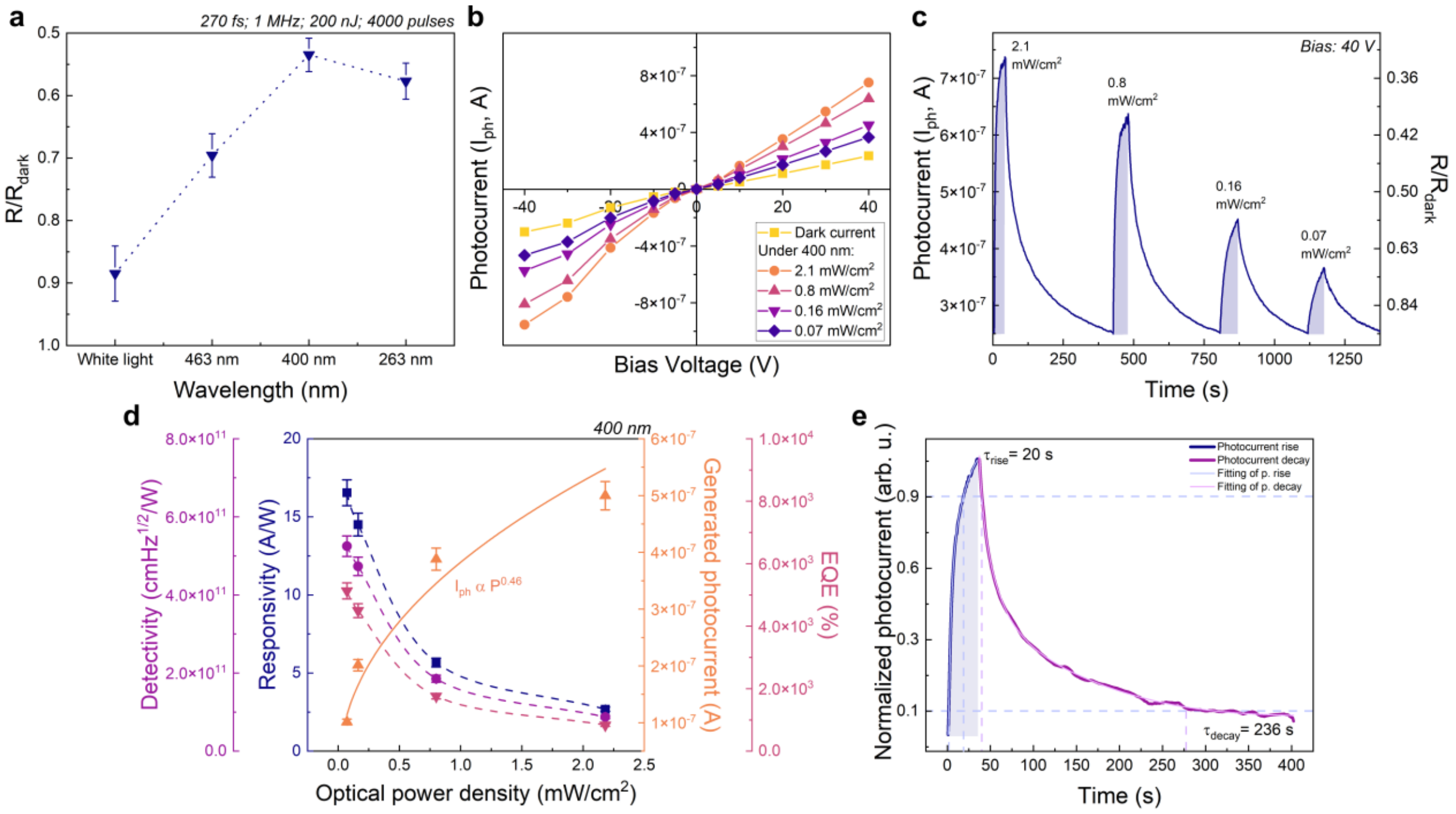

**Figure 3:** a) Spectral response of the device per unit of incident light power. b) I-V curve with and without illumination with various voltage biases (- 40 V to + 40 V). c) Typical temporal evolution with different optical power densities (0.07 to 2.1 mW/cm$^2$). e) Responsivity, detectivity, EQE (%), and generated photocurrent of Te/$TeO_2$-glass nanocomposite structure. e) Photocurrent rise and decay time for a line pattern. All measurements are performed under an open-air atmosphere at room temperature.

Table 1 summarizes the performance of various Te nanostructures and other nanostructures of photoconductive semiconductors such as CdS, CdSe, t-Se, and ZnO. While the size of the other Te devices is nano-/micro-scale, the laser-written patterns range from a few millimeters to a centimeter in length and hundreds of nanometers to a few micrometers in thickness. The responsivity and the detectivity vary with the optical power density. While this device exhibits comparable responsivity and detectivity with devices based on other manufacturing principles, there is still room to improve the time response. Let us investigate the stability of the device over time, which is an essential aspect of practical applications.

**Table 1:** Comparison of spectral range, responsivity, detectivity, the rise and decay time, and the stability of Te nanostructures and common photoconductive materials.

| Sample | Spectral range (nm) | Responsivity at RT (A/W) | Detectivity (Jones) | The rise and decay times | Stability |
|---|---|---|---|---|---|
| This work (10 mm patterns) | 263, 400, 463 nm and white light | 16.54 at 400 nm and 0.07 $mW/cm^2$ | $5.25\times10^{11}$ at 400 nm | 20 and 236 s | cycling more than 2 months |
| Te nanowires[48] | 633 nm | - | - | 40 and 40 s | cycling for 100 times |
| 2D Te nanoplates[49] | 413-550 nm | 389.5 at 473 nm and 76.2 $mW/cm^2$ | - | 4.4 and 2.8 s | bending for 100 times |
| 2D Te nanoflakes[31] | 1550 nm | 51.85 at 1550 nm and 0.51 $mW/mm^2$ | $1.88 \times 10^{10}$ | 19 and 21 μs | - |
| 2D Te nanoflakes[50] | 520, 1550 and 3390 nm | 383 at 520 nm and 1.6 nW | - | - | - |
| 2D Te nanoflakes[47] | 1400-2400 nm | 16 at 1700 nm | $2 \times 10^9$ | order of few seconds each | - |
| 2D Te nanosheets[30] | 350-400 nm | $13.4\times10^{-6}$ at 350 nm and 2.17 $mW/cm^2$ | $3.1\times 10^7$ | 54.5 and 70.2 ms | cycle stability for 10000s |
| Te nanoparticles in PMMA[22] | 310-2200 nm | $7.5\cdot10^{-8}$ at 400 nm and 4.2 $mW/cm^2$ | - | - | - |
| Te nanosheets and nanowires [51] | 830, 1310, 1550, 2000 nm and blackbody | 6650 at 1550 nm and 0.01 mW/mm$^2$ | 1.23 × 10$^{12}$ | 31.7 and 25.5 μs | +3 months |
| Te nanorods[23] | 300-785 nm | 6.1 at 0.94 mW/cm$^2$ | 1.2 × 10$^{11}$ | tens of seconds | ~30 days |
| CdS nanobelts[52] | 490 nm | 7.3 × 10$^4$ at 3 mW/cm$^2$ | - | ~20 μs | more than 73 hours |
| CdS nanorod[53] | 365, 420, 450, 500 nm | 1.23 × 10$^4$ at 450 nm and 0.5 mW/cm$^2$ | 2.8 × 10$^{11}$ | 0.82 and 0.84 s | - |
| CdSe nanocrystals[54] | 500-532 nm | 9.72 at 532 nm and 0.9 W/cm$^2$ | 6.9 × 10$^{10}$ | both below 2 μs | - |
| t-Se nanoparticles[55] | 300-700 nm | 19 × 10$^{-3}$ at 610 nm and 0.4 mW/cm$^2$ | - | 0.32 and 0.23 μs | - |
| ZnO nanowires[56] | 350-500 nm | 1109 at 356 nm | - | both ~tens of seconds | - |

**Figure 4** presents the photoconduction characteristics of the same device as in Figure 3, from the day of fabrication to a few months after. Figure 4a shows I-V curves obtained without and with illumination at 400 nm after one month. We notice that the dark current decreases from 0.24 µA to 0.09 µA with a bias of + 40 V after a month, and the current under illumination at 400 nm follows a similar trend. Figure 4b-c displays the temporal response upon various optical densities at 400 nm and corresponding responsivity, detectivity, quantum efficiency, and generated photocurrent values. R, D*, and EQE (%) values degrade over a month under illumination at 400 nm, while the exponent of generated photocurrent, α, stays the same. This behavior is attributed to trap states that are still present and that do not change over time. Finally, a repetitive measurement performed on the first day of the device fabrication is presented in Figure 4d, indicating the robustness and stability of the device. The amount of photo-response and dark resistivity after 10 hours of a cyclic test has not changed. Figure 4e-f shows the multiple on-off cycles after a few months of usage. The rise and decay time does not change after a month except for the photocurrent value. However, after two months, there is a noticeable baseline drift of the generated photocurrent and response times. There are a few possibilities to explain this dramatic change in the photo-response of the device over a few months. The first reason is the irreversible photo-induced oxidation upon UV illumination, observed previously in Te[4,7]. The oxidation of Te starts immediately after the first test of the device and forms $TeO_2$/Te/$TeO_2$-glass interface. However, the interface $TeO_2$/Te moves over time, as suggested by the dark current value being less than the value on the first day. Further illumination causes the oxidation of Te nanocrystals and nanoparticles, which results in the dark current approaching zero. To understand the degradation mechanism under UV irradiation, Raman spectra of pristine glass and the laser-written line patterns were collected over time with a Raman laser emitting at 445 nm.

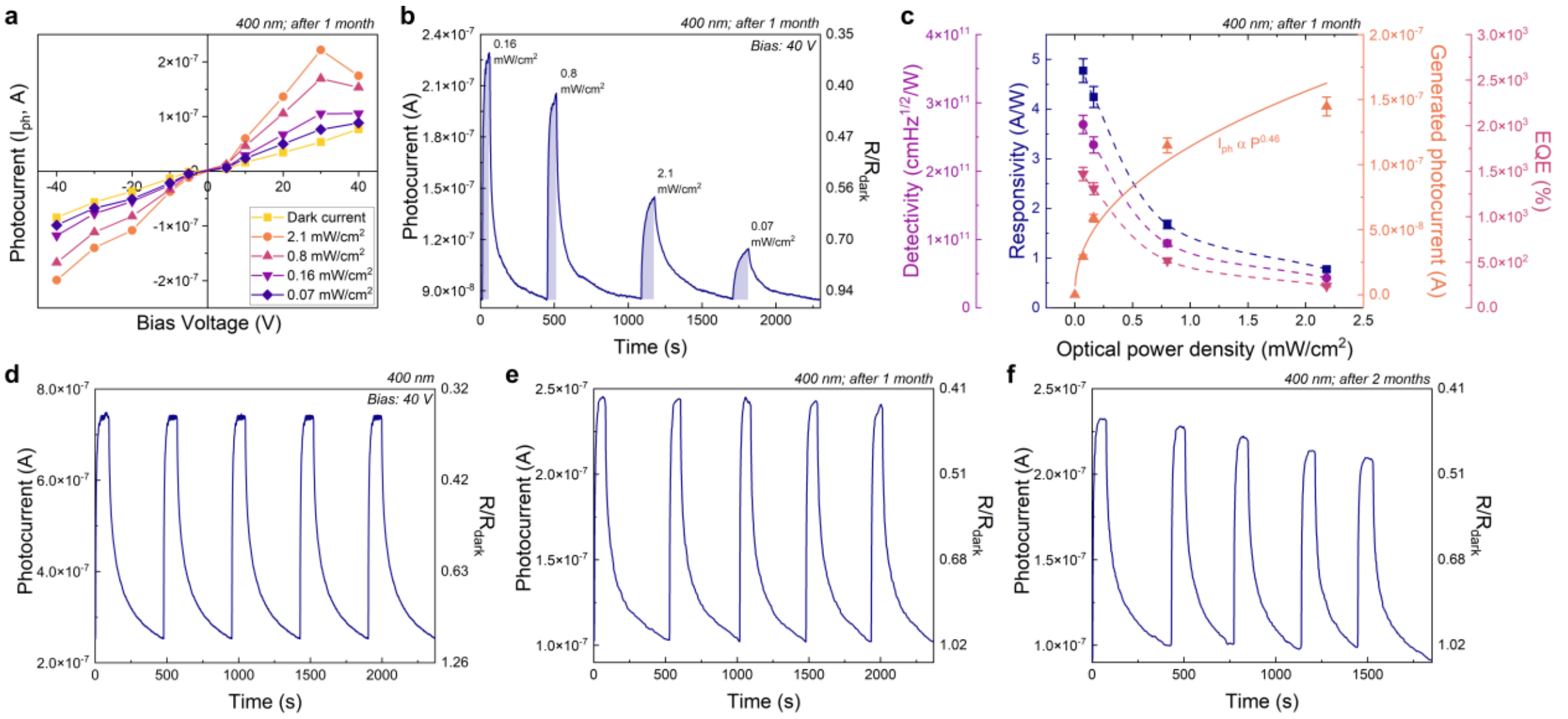

**Figure 4:** a) I-V curve without and with illumination at 400 nm for various voltage biases. The measurement was carried out a month after the fabrication of the device presented in Figure 3. b) Typical temporal evolution of photocurrent with different optical power densities (0.07 to 2.1 $mW/cm^2$) under 400 nm illumination one month after the fabrication of the device. c) Responsivity, detectivity, EQE (%), and generated photocurrent of Te/$TeO_2$-glass nanocomposite structure one month after the fabrication of the device. d-f) Repetitive temporal response at 400 nm with a flux of 2.1 $mW/cm^2$ observed on the day of the fabrication, after the first month and after the second months, respectively.

Raman spectra of the structural evolution of the tellurite glass under a Raman laser emitting at 445 nm are shown in **Figure 5a**. The glass network of the glass is composed of $TeO_4$ trigonal bipyramids (tbp), $TeO_{3+1}$ distorted trigonal bipyramids (d-tbp), $TeO_3$ trigonal pyramids (tp), $WO_4$, and $WO_6$ polyhedra. It results in the presence of Raman peaks located at around 355, 490, 610, 670, 720, 790, 860, and 920 $cm^{-1}$, and assigned to bending vibrations of W-O-W in $WO_6$ octahedra, a symmetrical stretching of Te-O-Te linkages, continuous network of $TeO_4$, antisymmetric stretching of Te-O-Te linkages consisting two inequivalent Te-O bonds, Te and NBO of $TeO_{3+1}$ and $TeO_3$, stretching of Te-$O^-$ in $TeO_{3+1}$ and $TeO_3$, stretching of W-O, W-$O^-$ and W=O bonds associated with $WO_4$ and $WO_6$ polyhedra, respectively[57,58]. There is no change in the peak intensity and ratio of the peaks in each spectrum collected at every 300 s. Hence, the glass substrate is not altered while characterizing the device.

Figure 5b-c displays the Raman spectra of the center of the laser-written pattern and the ratio of the Te versus $TeO_2$-glass. The characteristic vibration peaks of Te in the laser-modified zone are at 93, 118, 139, 170, and 260 $cm^{-1}$, corresponding to $E_1$, $A_1$, and $E_2$ modes, Te-Te homopolar bonds in amorphous Te (a-Te), and second-order spectra, respectively[7,21,59,60]. After irradiating for 300 s, the Raman spectra do not show any crystalline-$TeO_2$ peaks but rather a decrease in the intensity ratio of Te/$TeO_2$-glass and an increase in the intensity of the main glass bands. The photo-oxidation process further explains the longer decay time of the device after fabrication. Due to oxidation, the zones enriched with holes at the surface, the grain boundary, and the intragrain regions decrease. After two months of device usage, both rise and decay times decrease proportionally. Yet, time constants do not change, implying the presence of trap states as recombination centers. Fewer carriers are generating less photocurrent in each cycle, resulting in shorter recombination times. Transparent thin films of $SiO_2$ or $Si_3N_4$ can overcome the degradation of the patterns[27] by shielding the photoconductive layer from the environment. Another reason for the deterioration of the performances can be the degradation of the electrical contacts or the reaction between Te and the metal contacts (Au and Ag)[27]. Typical metal-contacted Te devices show a performance degradation over time and temperature (failure above 473K) due to the diffusion of metal atoms into the Te channels. A non-reactive interlayer, such as graphene[27], can be employed to prevent metal diffusion. In summary, preventative steps against detrimental effects such as the photo-oxidation of Te and the degradation of metallic contacts are necessary for better stability of the photoconductivity properties.

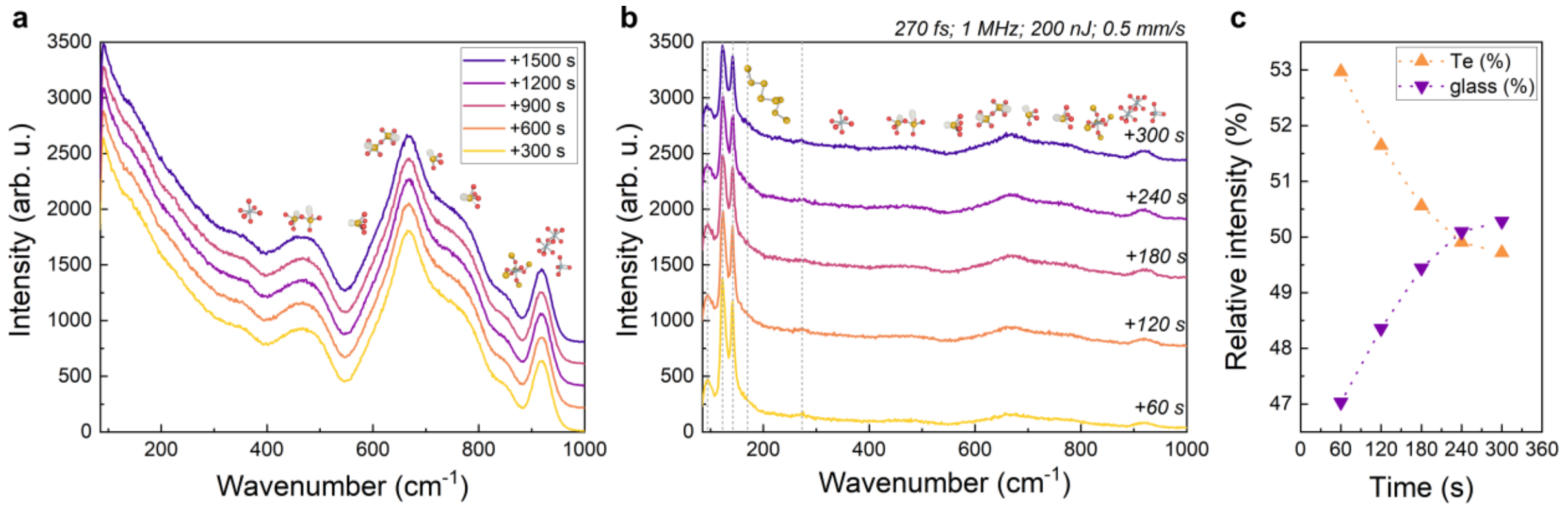

**Figure 5:** a) The Raman spectra of the pristine glass after irradiation with 445 nm-Raman laser for 1500 s. b) The Raman spectra of laser-irradiated line pattern after irradiation for 300 s. The femtosecond laser processing parameters are 200 nJ with 0.5 mm/s at 1 MHz (corresponding to an incoming pulse fluence of 262 J/mm$^2$). Note that Raman investigation was performed on the same day as the fabrication of the tested sample. c) The relative intensities of Te ($I_{93}+I_{122}+I_{141}+I_{170}+I_{260}$, (%)) and pristine glass peaks ($I_{356}+I_{470}+I_{610}+I_{670}+I_{720}+I_{790}+I_{860}+I_{920}$, (%)) at the laser-modified zone after irradiated with Raman laser for 300 s. The optical power density of the Raman laser is 115 mW/µm$^2$.

## 3. Summary and outlook

We observed a photo-response of a tellurium/tellurite glass (Te/$TeO_2$-glass) nanocomposite interface produced by a femtosecond laser direct-write process on a tellurite glass surface. After laser exposure, the measured resistivity is at least ten orders of magnitude lower than the one of the unmodified material and is comparable to the one value measured for polycrystalline tellurium. By scanning a femtosecond laser beam over the tellurite glass substrate, one can form write a conductive path between arbitrarily distant locations. Furthermore, we show that line patterns produced in this manner can have a highly reproducible and sensitive photo-response, from the near ultraviolet to the visible spectrum, stable over a few months. This manufacturing process demonstrates that one can turn an otherwise non-photo responsive substrate, in the present case tellurite glass, into a photoconductive one by exposing it to a femtosecond laser, and this, *without adding any material to the substrate*.

## Supporting Information

Supporting Information is available from the author.

**Acknowledgements**

The Galatea Lab is thankful to the sponsorship of Richemont International and to Tokyo University of Technology (TIT) for providing the tellurite glass substrates. The authors are thankful to Samuel Benketaf for contributing to the PCB design, Adrien Toros for wire bonding, and AQUA laboratory for providing a source measurement unit. The authors would like to thank Loïc Chautems for helping with the experimentation.

**Author's contributions**

In this paper, G.T. wrote the draft paper and performed most of the experiments presented in this paper. G.T., A.R., and Y.B. designed and developed the mask and device fabrication process. G.T. and A.R. performed resistivity measurements over temperature and spectral and temporal response of as-produced samples. G.T., A.R., and Y.B. interpreted the experimental results and analyzed the experimental data. T.K. produced the bulk material and measurement data of various bulk properties. Y.B. supervised the research. All the authors discussed and revised the paper.

**References**

[1] G. Torun, T. Kishi, Y. Bellouard, *Phys. Rev. Mater.* **2021**, *5*, 1.

[2] G. Torun, T. Kishi, D. Pugliese, D. Milanese, Y. Bellouard, *Adv. Mater.* **2023**, *2210446*, 1.

[3] J. Sunada, K. Oishi, A. Kasai, T. Kitahara, *Jpn. J. Appl. Phys.* **1982**, *21*, 1781.

[4] K. Oishi, K. Okamoto, J. Sunada, *Thin Solid Films* **1987**, *148*, 29.

[5] J. Sunada, K. Okamoto, K. Oishi, S. Shimazu, *Appl. Surf. Sci.* **1988**, *33–34*, 434.

[6] J. Sunada, Y. Hashimoto, K. Oishi, K. ich Fukuchi, *Phys. Lett. A* **1990**, *151*, 447.

[7] T. Vasileiadis, S. N. Yannopoulos, *J. Appl. Phys.* **2014**, *116*, DOI 10.1063/1.4894868.

[8] J. Sunada, K. Osada, T. Namekawa, K. Oishi, K. Fukuchi, *Phys. Lett. A* **1990**, *146*, 85.

[9] M. Palomba, U. Coscia, G. Carotenuto, S. De Nicola, G. Ambrosone, *Phys. Status Solidi Curr. Top. Solid State Phys.* **2015**, *12*, 1317.

[10] D. Tsiulyanu, S. Marian, V. Miron, H. D. Liess, *Sensors Actuators, B Chem.* **2001**, *73*, 35.

[11] D. Tsiulyanu, *Beilstein J. Nanotechnol.* **2020**, *11*, 1010.

[12] F. Arab, M. Mousavi-Kamazani, M. Salavati-Niasari, *RSC Adv.* **2016**, *6*, 71472.

[13] R. N. Hampton, W. Hong, G. A. Saunders, R. A. El-Mallawany, *J. Non. Cryst. Solids*

**1987**, *94*, 307.

[14] M. Çelikbilek, *Ph.D. Thesis, Istanbul Tech. Univ.* **2013**.

[15] R. W. McKay, W. E. Gravelle, *Can. J. Phys.* **1961**, *39*, 534.

[16] A. Nussbaum, *Phys. Rev.* **1954**, *94*, 337.

[17] M. A. Dinno, M. Schwartz, B. Giammara, *J. Appl. Phys.* **1974**, *45*, 3328.

[18] G. Fischer, G. K. White, S. B. Woods, *Phys. Rev.* **1957**, *106*, 480.

[19] X. Zhang, J. Jiang, A. A. Suleiman, B. Jin, X. Hu, X. Zhou, T. Zhai, *Adv. Funct. Mater.* **2019**, *29*, DOI 10.1002/adfm.201906585.

[20] X. Zhao, J. Shi, Q. Yin, Z. Dong, Y. Zhang, L. Kang, Q. Yu, C. Chen, J. Li, X. Liu, K. Zhang, *iScience* **2022**, *25*, 103594.

[21] C. Marini, D. Chermisi, M. Lavagnini, D. Di Castro, C. Petrillo, L. Degiorgi, S. Scandolo, P. Postorino, *Phys. Rev. B - Condens. Matter Mater. Phys.* **2012**, *86*, 1.

[22] U. Coscia, G. Ambrosone, M. Palomba, S. Binetti, A. Le Donne, D. Siliqi, G. Carotenuto, *Appl. Surf. Sci.* **2018**, *457*, 229.

[23] Q. Xiao, X. Li, Z. Zhang, C. Hu, G. Dun, B. Sun, Y. Peng, Q. Wang, Z. Zheng, H. Zhang, *Adv. Electron. Mater.* **2020**, *6*, DOI 10.1002/aelm.202000240.

[24] C. Ma, J. Yan, Y. Huang, C. Wang, G. Yang, *Sci. Adv.* **2018**, *4*, 1.

[25] P. Makuła, M. Pacia, W. Macyk, *J. Phys. Chem. Lett.* **2018**, *9*, 6814.

[26] D. Tsiulyanu, O. Mocreac, *Sensors Actuators, B Chem.* **2013**, *177*, 1128.

[27] C. Zhao, L. Hurtado, A. Javey, *Appl. Phys. Lett.* **2020**, *117*, DOI 10.1063/5.0018045.

[28] V. Selamneni, T. Akshaya, V. Adepu, P. Sahatiya, *Nanotechnology* **2021**, *32*, DOI 10.1088/1361-6528/ac19d8.

[29] M. Peng, Y. Yu, Z. Wang, X. Fu, Y. Gu, Y. Wang, K. Zhang, Z. Zhang, M. Huang, Z. Cui, F. Zhong, P. Wu, J. Ye, T. Xu, Q. Li, P. Wang, F. Yue, F. Wu, J. Dai, C. Chen, W. Hu, *ACS Photonics* **2022**, *9*, 1775.

[30] Z. Xie, C. Xing, W. Huang, T. Fan, Z. Li, J. Zhao, Y. Xiang, Z. Guo, J. Li, Z. Yang, B. Dong, J. Qu, D. Fan, H. Zhang, *Adv. Funct. Mater.* **2018**, *28*, 1.

[31] Y. Yan, K. Xia, W. Gan, K. Yang, G. Li, X. Tang, L. Li, C. Zhang, G. T. Fei, H. Li, *Nanoscale* **2022**, *14*, 13187.

[32] Q. Zhao, W. Wang, F. Carrascoso-Plana, W. Jie, T. Wang, A. Castellanos-Gomez, R. Frisenda, *Mater. Horizons* **2020**, *7*, 252.

[33] Q. Zhang, J. Jie, S. Diao, Z. Shao, Q. Zhang, L. Wang, W. Deng, W. Hu, X. Hui, X. Yuan, S.-T. Lee, *ACS Nano* **2015**, *9*, 1561.

[34] H. G. Junginger, *Solid State Commun.* **1967**, *5*, 509.

[35] A. Kramer, M. L. Van de Put, C. L. Hinkle, W. G. Vandenberghe, *npj 2D Mater. Appl.* **2020**, *4*, 1.
[36] M. Razeghi, A. Rogalski, *J. Appl. Phys.* **1996**, *79*, 7433.
[37] F. Alema, B. Hertog, O. Ledyaev, D. Volovik, R. Miller, A. Osinsky, S. Bakhshi, W. V. Schoenfeld, *Sensors Actuators, A Phys.* **2016**, *249*, 263.
[38] K. Okuyama, Y. Kumagai, *Jpn. J. Appl. Phys.* **1973**, *12*, 1884.
[39] J. Miao, C. Wang, *Nano Res.* **2021**, *14*, 1878.
[40] P. Bhaskar, A. W. Achtstein, M. J. W. Vermeulen, L. D. A. Siebbeles, *J. Phys. Chem. C* **2019**, *123*, 841.
[41] V. D. Popovych, M. Bester, *J. Appl. Phys.* **2012**, *112*, 1.
[42] D. Guo, Y. Su, H. Shi, P. Li, N. Zhao, J. Ye, S. Wang, A. Liu, Z. Chen, C. Li, W. Tang, *ACS Nano* **2018**, *12*, 12827.
[43] Y. Zhang, D. J. Hellebusch, N. D. Bronstein, C. Ko, D. F. Ogletree, M. Salmeron, A. P. Alivisatos, *Nat. Commun.* **2016**, *7*, DOI 10.1038/ncomms11924.
[44] S. Jeon, S. E. Ahn, I. Song, C. J. Kim, U. I. Chung, E. Lee, I. Yoo, A. Nathan, S. Lee, J. Robertson, K. Kim, *Nat. Mater.* **2012**, *11*, 301.
[45] A. George, M. V. Fistul, M. Gruenewald, D. Kaiser, T. Lehnert, R. Mupparapu, C. Neumann, U. Hübner, M. Schaal, N. Masurkar, L. M. R. Arava, I. Staude, U. Kaiser, T. Fritz, A. Turchanin, *npj 2D Mater. Appl.* **2021**, *5*, DOI 10.1038/s41699-020-00182-0.
[46] E. Arslan, S. Bütün, S. B. Lisesivdin, M. Kasap, S. Ozcelik, E. Ozbay, *J. Appl. Phys.* **2008**, *103*, DOI 10.1063/1.2921832.
[47] M. Amani, C. Tan, G. Zhang, C. Zhao, J. Bullock, X. Song, H. Kim, V. R. Shrestha, Y. Gao, K. B. Crozier, M. Scott, A. Javey, *ACS Nano* **2018**, *12*, 7253.
[48] Y. Wang, Z. Tang, P. Podsiadlo, Y. Elkasabi, J. Lahann, N. A. Kotov, *Adv. Mater.* **2006**, *18*, 518.
[49] Q. Wang, M. Safdar, K. Xu, M. Mirza, Z. Wang, J. He, *ACS Nano* **2014**, *8*, 7497.
[50] C. Shen, Y. Liu, J. Wu, C. Xu, D. Cui, Z. Li, Q. Liu, Y. Li, Y. Wang, X. Cao, H. Kumazoe, F. Shimojo, A. Krishnamoorthy, R. K. Kalia, A. Nakano, P. D. Vashishta, M. R. Amer, A. N. Abbas, H. Wang, W. Wu, C. Zhou, *ACS Nano* **2020**, *14*, 303.
[51] M. Peng, R. Xie, Z. Wang, P. Wang, F. Wang, H. Ge, Y. Wang, F. Zhong, P. Wu, J. Ye, Q. Li, L. Zhang, X. Ge, Y. Ye, Y. Lei, W. Jiang, Z. Hu, F. Wu, X. Zhou, J. Miao, J. Wang, H. Yan, C. Shan, J. Dai, C. Chen, X. Chen, W. Lu, W. Hu, *Sci. Adv.* **2021**, *7*, 1.
[52] L. Li, P. C. Wu, X. S. Fang, T. Y. Zhai, L. Dai, M. Y. Liao, Y. Koide, H. Q. Wang, Y. Bando, D. Golberg, *Adv. Mater.* **2010**, *22*, 3161.

[53] W. Zhao, L. Liu, M. Xu, X. Wang, T. Zhang, Y. Wang, Z. Zhang, S. Qin, Z. Liu, *Adv. Opt. Mater.* **2017**, *5*, 1.

[54] W. Xing, S.-C. Kung, W. E. van der Veer, W. Yan, T. Ayvazian, J. Y. Kim, R. M. Penner, *ACS Nano* **2012**, *6*, 5627.

[55] K. Hu, H. Chen, M. Jiang, F. Teng, L. Zheng, X. Fang, *Adv. Funct. Mater.* **2016**, *26*, 6641.

[56] K. Keramatnejad, F. Khorramshahi, S. Khatami, E. Asl-Soleimani, *Opt. Quantum Electron.* **2015**, *47*, 1739.

[57] T. Kosuge, Y. Benino, V. Dimitrov, R. Sato, T. Komatsu, *J. Non. Cryst. Solids* **1998**, *242*, 154.

[58] M. Çelikbilek Ersundu, A. E. Ersundu, M. I. Sayyed, G. Lakshminarayana, S. Aydin, *J. Alloys Compd.* **2017**, *714*, 278.

[59] Y. H. Cheng, S. W. Teitelbaum, F. Y. Gao, K. A. Nelson, *Phys. Rev. B* **2018**, *98*, DOI 10.1103/PhysRevB.98.134112.

[60] R. T. Ananth Kumar, H. A. Mousa, P. Chithra Lekha, S. T. Mahmoud, N. Qamhieh, *J. Phys. Conf. Ser.* **2017**, *869*, DOI 10.1088/1742-6596/869/1/012018.